# Testing the Influence of Temperature on Mass at High Temperatures

M. Tajmar[*], G. Hentsch

Institute of Aerospace Engineering, Technische Universität Dresden, 01307 Dresden, Germany

T. Hutsch

Fraunhofer Institute for Manufacturing Technology and Advanced Materials IFAM, Branch Lab Dresden, 01277 Dresden, Germany

**Abstract**

Special relativity predicts a very small influence of temperature on mass of around $\frac{\Delta m}{m} \approx 10^{-14}$. More than 100 years ago, experiments were performed that revealed a limit of <$10^{-8}$ for changes of a few degree at room temperature. A similar limit was obtained with a magnetic suspension balance at cryogenic temperatures. Recently, some measurements claim to have seen a negative dependence at the $10^{-6}$ level for a variety of metal samples at room temperature, which would be of interest e.g. to explain variations in the measurement of the gravitational constant. We have performed measurements with an analytical balance in vacuum and with a commercial thermogravimetric balance in argon and obtained a 3σ limit of $< 1.8 \times 10^{-8}$ for both metallic (Cu, Pt) and non-metallic ($Al_2O_3$) samples ranging from room temperature to above 1000°C. This extends previous measurements to the high-temperature regime and rules out all claimed anomalies by more than two orders of magnitude.

## 1. Introduction

Does the temperature of a body influence its mass? According to $E = mc^2$ it will and we can approximate the change in mass of a single particle due to a change in temperature by using the Dulong-Petit law for thermal energy [1] as

$$\Delta m = \frac{3k_B \Delta T}{c^2} \tag{1}$$

where $k_B$ is Boltzmann's constant. We may also normalize this equation with respect to the initial particle mass $m_0$ of a sample body and its temperature change leading to

$$\gamma = \frac{\Delta m}{m_0 \Delta T} = \frac{3k_B}{m_0 c^2} \tag{2}$$

This effect of mass change with temperature should therefore be dominant for lighter atoms and gives $\gamma_H = 2.8 \times 10^{-13} K^{-1}$ or $\gamma_{Cu} = 4.4 \times 10^{-15} K^{-1}$ for the case of hydrogen and copper respectively. These numbers seem exceedingly small to be measured directly. The first attempt was done by Poynting and Philipps [2] some 100 years ago claiming $\gamma < 2 \times 10^{-10} K^{-1}$, which is well above our estimate. This value seems outstanding for the time of its measurement, however there are concerns that this value is not correct. They used a knife-edge balance in vacuum and heated or cooled a brass mass only with radiative heat transfer from a hot water steam or liquid-air filled jacket next to their test mass without measuring the temperature directly. Shortly afterwards, Southerns [3] reported a value of $\gamma < 10^{-8} K^{-1}$ for a copper sample in a temperature-controlled calorimeter also on a vacuum knife-edge balance with much better control, however, only within a temperature range of 11-32 °C.

---

[*] Corresponding Author, EMail: martin.tajmar@tu-dresden.de

A decade later, Shaw [4] tried to detect any influence of temperature on gravitational attraction by building a Cavendish-type gravitational-constant-type setup with a torsion pendulum inside a vacuum tube with large lead masses on the outside that could be heated up to 250 °C. This is complementary to the previous attempts that concentrated on measuring the influence of the smaller mass with their weight measurement using the Earth as the larger mass. The initial experiment was improved by Shaw and Davy [5] reporting $\gamma < 2 \times 10^{-6} K^{-1}$.

Some 80 years later, Dmitriev [6–8] re-analyzed the Shaw and Davy experiment by looking for a connection between elastic forces and gravitational acceleration that should lead to a much larger than classically expected mass-temperature relationship. He performed weight experiments with a variety of mostly metallic samples that were heated up by an electrical or chemical heater or even by ultrasonic means. Although his temperature changes were rather small at the degree range, and without simultaneously measuring weight and temperature of his sample, he claimed consistent $\gamma$ values in the negative $10^{-6}$ range, meaning that he observed a weight loss at higher temperatures. Similarly, Liangzao et al [9] also reported a negative $\gamma$ in the $10^{-6}$ range for metallic samples within a temperature range of 100-600°C. However, the experiment seems too simple with weight measurements of samples out of a heater without any control of the environment.

A real temperature-mass influence above classic predictions would be of course of great interest and may even resolve some of the discrepancies of measuring the gravitational constant [10], which has the biggest uncertainty of all fundamental constants. Tajmar et al [11] reported a limit of $\gamma < 2 \times 10^{-8} K^{-1}$ for copper using a high-vacuum magnetic suspension balance with simultaneous measurement of the sample's weight and its temperature for a range of -193 to -43°C.

In this work, we want to extend our search for any anomalies in metallic and non-metallic samples towards high temperatures close to 1000 °C, which is some 400 °C higher than previous tests. We performed measurements using an analytical balance in vacuum as well as using a commercial thermogravimetric balance in argon atmosphere for both metallic and non-metallic samples, which rules out all claimed anomalies by more than two orders of magnitude. A summary of all past and present results is shown in **Table 1**. The paper is structured as follows: First, the setup and experiments are described for the analytical vacuum balance. Next, a similar presentation is done for the commercial thermogravimetric balance, where we included a discussion on the buoyancy correction due to the argon environment. The paper then closes with a conclusion, which summarizes the main findings.

**2. Experiments**

We performed our experiments with two setups that are complementary and allow to compare results for similar sample materials:

- Analytical balance in vacuum with sample heating through radiative heat transfer from surrounding heater inside a thermal shield/cooling assembly
- Commercial thermogravimetric balance (Netzsch STA 499C Jupiter) with dynamic argon gas flow atmosphere

Both setups allow to measure the sample's temperature directly for a reliable mass-temperature assessment.

*2.1 Setup 1: Analytical Vacuum Balance*

For this measurement, we modified a commercial analytical balance VWR LA614i for operation in vacuum by removing all plastic parts separating the electronics and display unit with a cable that can be passed through a vacuum feedthrough flange. It featured a maximum load of 610 g and a precision of 0.1 mg. In order to eliminate thermal expansion introduced drifts due to a change in the centre of mass, we decided to put the test sample below the balance using a yoke frame as shown in **Fig. 1** (this was necessary because the balance did not feature a hook). The sample used was a cylindrical piece of copper machined in our workshop with a purity of >99.9% and a mass of 320.19 g, which was clamped to an alumina rod with multiple capillaries using a thin tungsten wire. A K-type thermocouple was directly attached to the sample and connected through the alumina capillaries back to the yoke. We then used Galinstan liquid metal contacts to pass the signal to a conventional thermocouple cable without influencing the weight measurements by stiff connections. Temperature calibration showed that this setup gave reliable readings within the typical K-type thermocouple limits. A Labjack T7 was used to read the temperature.

The sample was placed inside a ceramic heater with tungsten heater wires. The whole assembly was thermally isolated using Zircar ZAL-15 as a low thermal conductive shield for high temperatures and a water cooled surface on the outside. This protected the balance from the high temperatures of the heater that could lead to large thermal drifts. A constant heater power of 200 W was used to achieve an equilibrium heater temperature of around 700°C. The steps to perform the measurement are summarized as follows:

1. Dismount the cooler and thermo-shield top cap and put the alumina capillary through them
2. Attach sample to wire
3. Mount wire assembly to balance yoke and re-attached cooler and thermo-shield caps
4. Pump down chamber and wait until vacuum level is stabilized
5. Turn on cooler and start heating
6. Start balance measurement

All experiments were done in vacuum using a large vacuum chamber and an Edwards XDS35i scroll pump. Key components are shown in **Fig. 2**. Due to a small leak in the water cooler, we only achieved pressures in the mbar range. The sample was therefore only heated via radiative heat transfer. Three measurements are shown in **Fig. 3**, illustrating the heating up and cooling down phase with the recorded mass changes. Some large weight variations usually happened towards the maximum reachable sample temperature around 600°C, which we believe is due to outgassing of the heater and different thermal expansion coefficients from the alumina tube, the tungsten wire and the copper test mass that causes sudden mechanical twists and a rearrangement of the sample's equilibrium point. We therefore only included regions with low noise in our analysis as indicated by the coloured regions in the graphs. The results of our linear regression analysis are shown in **Table 2**. We achieved a $\gamma$ value in the $10^{-8}$ range with a tendency towards a negative correlation and coefficients of determination ($R^2$) in the range of 0.5-0.8. This weak correlation may very well be setup-related and therefore we decided to compare our results with a commercial thermogravimetric balance and average all measurements to achieve better statistics.

*2.2 Setup 2: Commercial Thermogravimetric Balance*

Thermogravimetric analysis (TGA) is a well-established method to measure mass changes with respect to changes in atmosphere and temperature. This is an important tool for a variety of investigations like material outgassing/absorption, phase-transitions, or chemical solid-gas reactions. We may use such a tool as well for our own analysis, if we use sample materials that do not react with the instrument's atmosphere within our temperature range. In fact, thermal drifts of the measured mass are well-known since the early stages of TGA balances [12]. Usually, two methods are used to reduce drifts and facility-induced side-effects:

1. Performing a measurement with and without sample to eliminate effects from the balance itself
2. Subtracting buoyancy-related effects using the ambient gas parameters as well as the sample's volume

We will follow the same procedures for our test. The commercial thermogravimetric balance was a Netzsch STA 449C Jupiter. The sample was put into an alumina cup holder, which was surrounded by a non-contact heater and connected to the balance cell by a long ceramic rod at the bottom. The actual temperature was measured by an S-type thermocouple directly below the sample. The balance could operate under dynamic gas flow, static atmosphere or vacuum conditions. Normally the best results for this kind of test can be achieved by operating in vacuum. However, because the STA 449C is normally used for calorimetric measurements like evaporation processes, it is optimized for measurements under dynamic gas flow, which also provides a good thermal contact. We therefore decided to use this mode of operation with inert argon gas. A summary of the samples used is shown in **Table 3**. In addition to copper, we also used platinum and alumina ($Al_2O_3$) as a non-metallic sample. The sample's mass was now much smaller at around 4 g, however, the balance's resolution was much better too at 0.5 µg.

The buoyancy correction is done by subtracting the measured mass change with the following equation:

$$\Delta m = \rho_0 V_0 - \frac{p_0 V_0}{R_s T} \cdot (1 + 3\alpha \Delta T) \tag{3}$$

where the first term describes the mass of the displaced gas at normal conditions with density $\rho_0$ and sample volume $V_0$, and the second term the change in buoyancy due to the variation of the gas density and volume from thermal expansion using the linear thermal expansion coefficient $\alpha$ as well as the specific gas constant $R_s$. The values used for the gas can be found in **Table 3** and **4**.

The heating rate was set to 10 K/min for all experimental runs. Because of the high reactivity to remaining oxygen in the atmosphere of copper, the target temperature was lowered to 350 °C for this sample, to protect the sample for unwanted oxidation side effects. For platinum and alumina, the target temperatures were set to 1000 °C to extend our analysis temperature range as much as possible. All correction and sample runs were performed twice.

The measured mass change over the temperature profile, already taking into account the correction measurement without the sample, as well as the calculated mass change due to buoyancy in **Equ. (3)** are shown in **Fig. 4**. Both matches quite well indicating that our calculations are correct. This buoyancy correction is now subtracted from the measured mass change and again suitable regions for our linear regression analysis were identified as shown in **Fig. 5**.

The results are again summarized in **Table 1**. Here, the coefficients of determination ($R^2$) are low indicating that there is no linear correlation between mass changes and temperature changes. Our $\gamma$ factors are all in the $10^{-9}$ or low $10^{-8}$ range similar to the results from Setup 1. This confirms our suspicion, that the weak correlation in Setup 1 was only setup induced.

## 3. Conclusion

No influence of temperature on mass was found using both a standard analytical balance in vacuum as well as a commercial thermogravimetric balance in argon atmosphere for both metallic (Cu, Pt) and non-metallic ($Al_2O_3$) samples. Our analysis confirms the old limit range set by Southerns [3] more than 100 years ago but extends it from room temperature for the first time up to 1065 °C. We may use all obtained $\gamma$ mean values to compute a 3σ mean experimental limit of $\gamma_{all} < 1.8 \times 10^{-8}$. This rules out the anomalies claimed by Dmitriev [6–8] and Liangzao [9] by more than two orders of magnitude.

**Acknowledgement**

We gratefully acknowledge the support by the German National Space Agency DLR (Deutsches Zentrum fuer Luft- und Raumfahrttechnik) by funding from the Federal Ministry of Economic Affairs and Energy (BMWi) by approval from German Parliament (50RS1704).

| Reference | Material | Method | Temperature Range | $|\gamma|$ [1/K] | Comment |
|---|---|---|---|---|---|
| Poynting and Phillips [2] | Brass | Knife-edge Balance in Vacuum, one mass at RT, one subjected to $\Delta T$ via radiation | -186°C-100°C | $<2\times10^{-10}$ | No direct $T_{sample}$ measurement |
| Southerns [3] | Copper | Knife-edge Balance in Vacuum, one mass at RT and one subjected to $\Delta T$ by heater on balance | 11°C-32°C | $<10^{-8}$ | |
| Shaw and Davy [5] | Lead | Vacuum Cavendish-Balance - Heating up larger mass | 20°C-250°C | $<2\times10^{-6}$ | |
| Dmitriev [6–8] | Al, Ti, Brass, Cu, Pb, Steel, PZT | Chemical Heating, Electric Resistance, Ultrasonic Heating | RT, $\Delta T$ up to 10°C | $1\text{-}10\times10^{-6}$ | No direct $T_{sample}$ measurement |
| Liangzao et al [9] | Au, Ag, Cu, Fe, Ni, Al | Remove sample from heater and put on balance | 100°C-600°C | $\approx 1\times10^{-6}$ | Too simple measurement |
| Tajmar et al [11] | Copper | Vacuum Magnetic Suspension Balance | -193°C ~ -43°C | $<2\times10^{-8}$ | |
| This work | Cu, Pt, $Al_2O_3$ | Balance in Vacuum or Argon Atmosphere | 33°C-1060°C | $<1.8\times10^{-8}$ | |

**Table 1** Summary of Temperature-Mass Measurements

| Material | Atmosphere | $T_{min}$ [°C] | $T_{max}$ [°C] | $\Delta T$ [K] | Mass [g] | $\gamma$ [1/K] | $R^2$ |
|---|---|---|---|---|---|---|---|
| Copper | Vacuum (8 mbar) | 78 | 361 | 283 | 320.19 | $-2.1\times10^{-8} \pm 5.1\times10^{-11}$ | 0.81 |
|  |  | 79 | 339 | 260 |  | $1.6\times10^{-8} \pm 4.2\times10^{-11}$ | 0.47 |
|  | Vacuum (16 mbar) | 33 | 396 | 363 |  | $2.1\times10^{-9} \pm 2.4\times10^{-11}$ | 0.03 |
|  | Vacuum (5 mbar) | 213 | 613 | 400 |  | $-1.0\times10^{-8} \pm 4.2\times10^{-11}$ | 0.63 |
|  |  | 267 | 525 | 258 |  | $-2.0\times10^{-8} \pm 3.7\times10^{-11}$ | 0.85 |
| Copper | Argon | 119 | 393 | 274 | 4.85 | $4.9\times10^{-9} \pm 1.0\times10^{-9}$ | 0.01 |
|  |  |  |  |  |  | $4.0\times10^{-8} \pm 3.0\times10^{-9}$ | 0.05 |
| Platinum | Argon | 115 | 1065 | 950 | 4.16 | $1.3\times10^{-9} \pm 4.7\times10^{-11}$ | 0.07 |
|  |  |  |  |  |  | $7.8\times10^{-9} \pm 1.4\times10^{-11}$ | 0.97 |
| $Al_2O_3$ | Argon | 127 | 1060 | 933 | 3.56 | $7.5\times10^{-9} \pm 1.2\times10^{-10}$ | 0.27 |
|  |  |  |  |  |  | $1.0\times10^{-8} \pm 1.6\times10^{-10}$ | 0.27 |

**Table 2** Measurements Summary (Upper Half with Setup 1/VWR 614i, Lower Half with Setup 2/ Netzsch STA 449C Jupiter)

| Material | Shape | Dimensions [mm] | Volume [m³] | Mass [mg] | Thermal Expansion Coeff. $\alpha$ [$K^{-1}$] | Target Temperature [°C] |
|---|---|---|---|---|---|---|
| Copper | Cube | 8.3×7.93×8.63 | $5.21\cdot10^{-7}$ | 4845.6 | $16.5\times10^{-6}$ | 350 |
| Platinum (90Pt10Rh) | Batch of wire | Ø0.5×1074.1 | $2.11\cdot10^{-7}$ | 4158.1 | $9\times10^{-6}$ | 1000 |
| Aluminium Oxide (99.9%) | Cylinder | Ø10×11.78 | $9.25\cdot10^{-7}$ | 3561.3 | $8\times10^{-6}$ | 1000 |

**Table 3** Samples measured with Netzsch STA 449C Jupiter Balance

| Property | $p_0$ [Pa] | $\rho_0$ [Kg/m³] | $R_s$ [$J\cdot Kg^{-1}\cdot K^{-1}$] |
|---|---|---|---|
| Value | 101325 | 1.642 | 208.1 |

**Table 4** Argon Properties used for Buoyancy Compensation

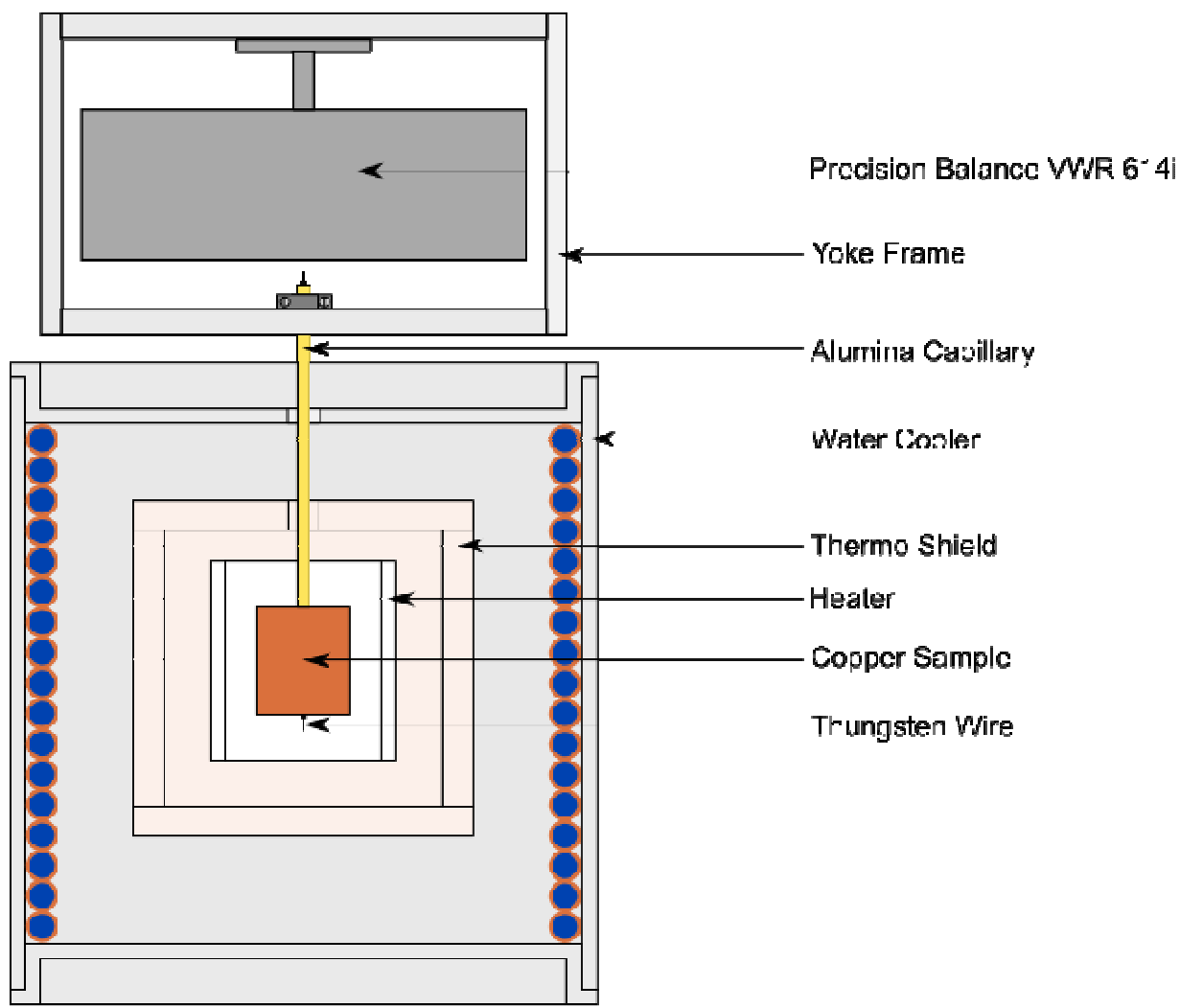

**Fig. 1**   Setup 1 – VWR Balance

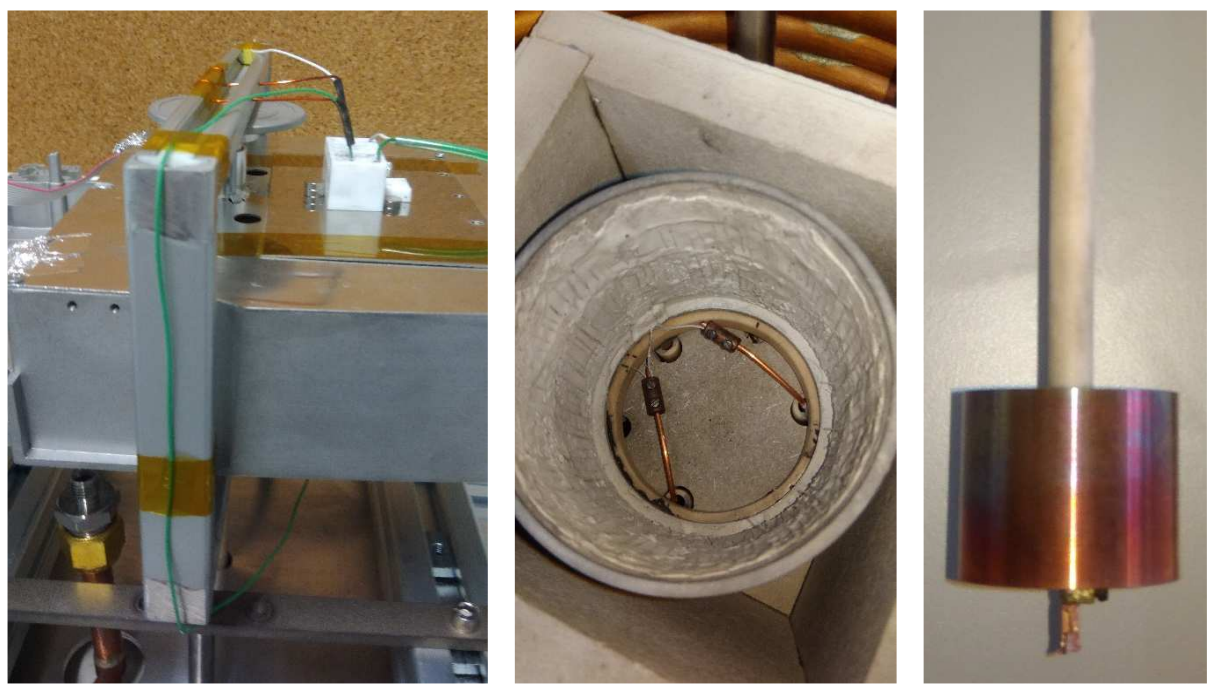

**Fig. 2** Pictures of Key Components from Setup 1 (From Left to Right): VWR Balance with Yoke Frame and Liquid-Metal Connection for K-Thermocouple, Heater inside Thermal Shield, Copper Test Mass on Alumina Capillary and Thermocouple on the Bottom

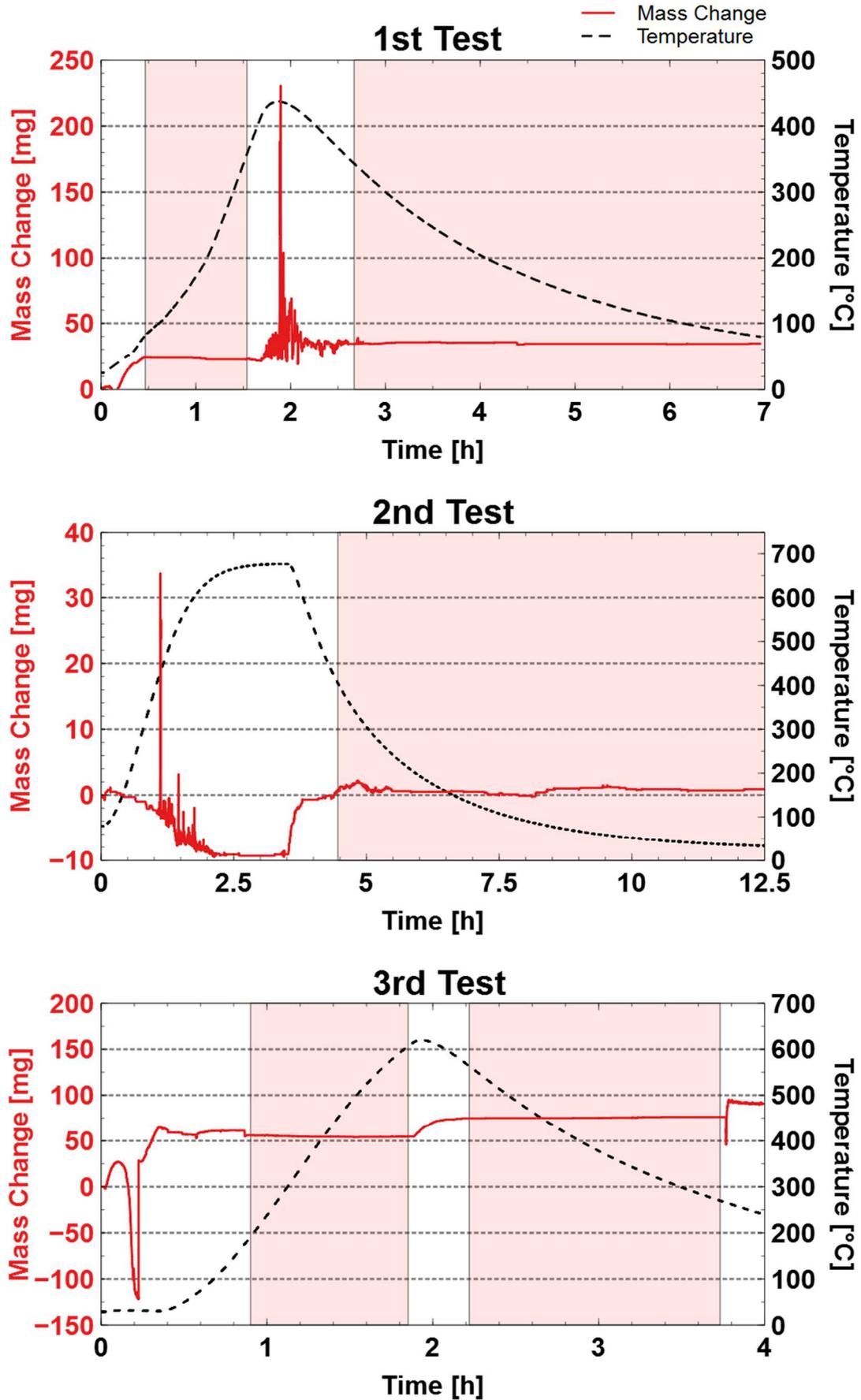

**Fig. 3**  Mass Change Measurement with Setup 1 (VWR Balance) – Colored Regions used for Analysis

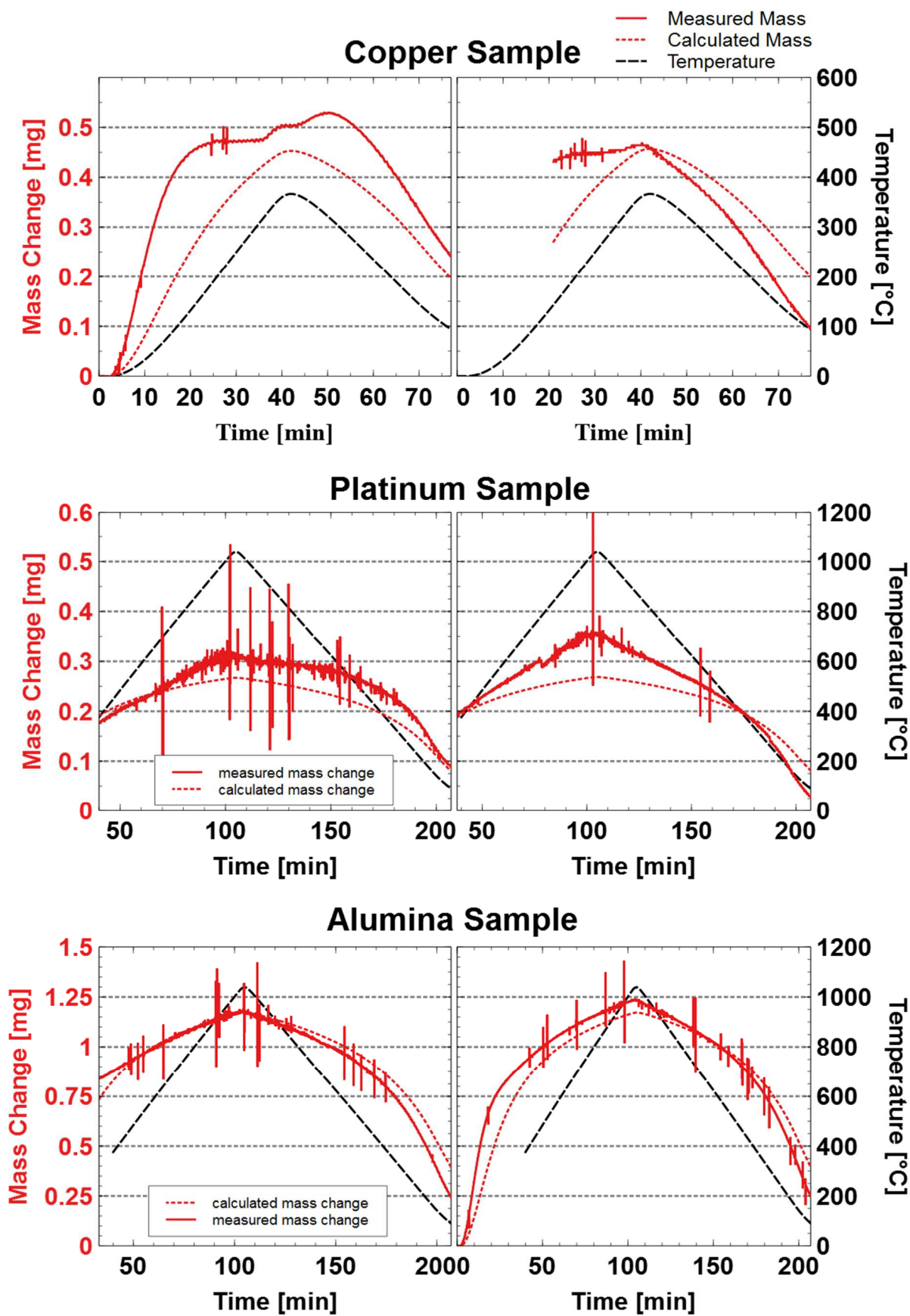

**Fig. 4** Two Calculated versus Measured Mass Changes for Each Sample in Setup 2 (Netzsch Balance)

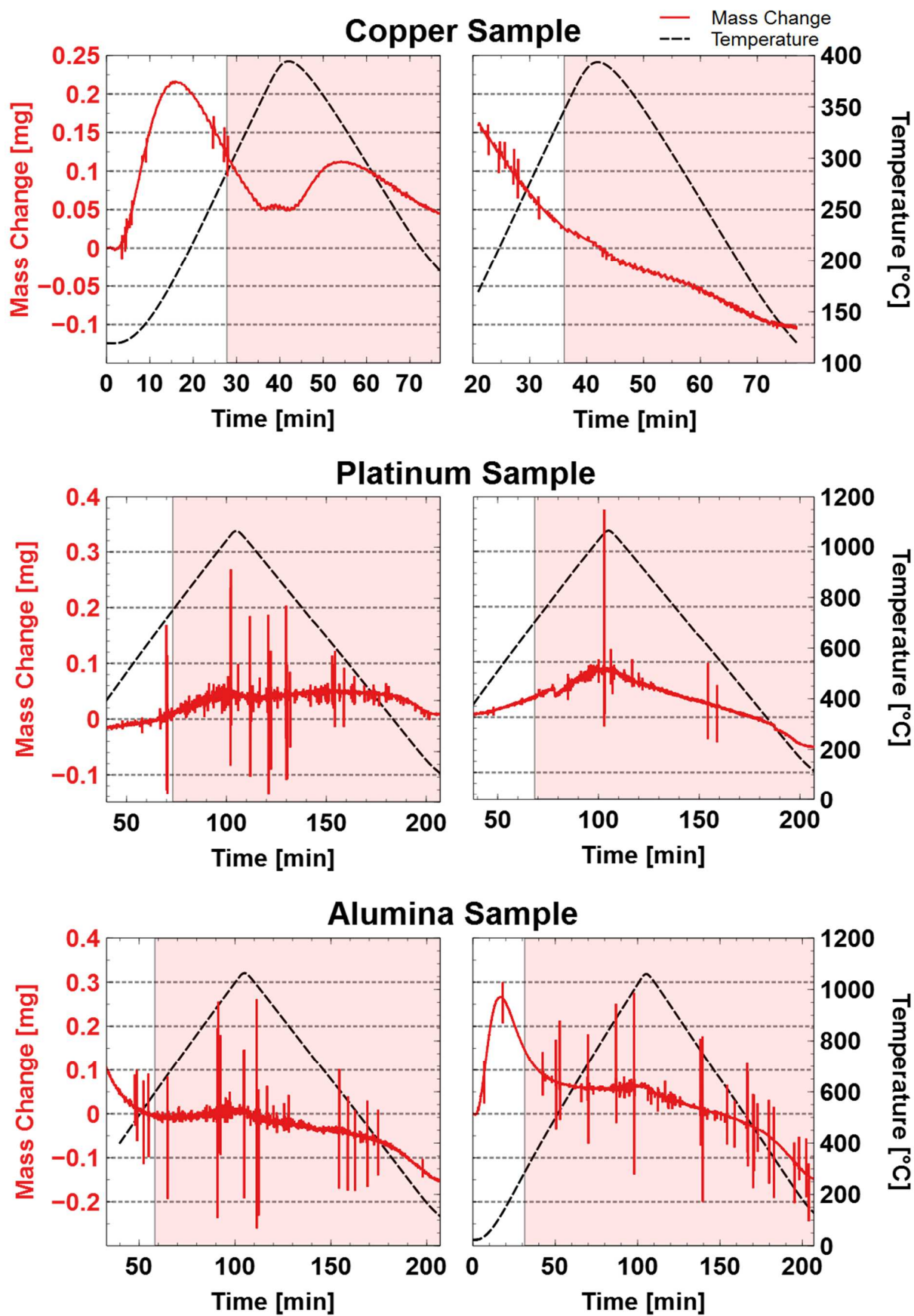

**Fig. 5** Two Mass Change Measurements including Buoyancy Correction for Each Sample with Setup 2 (Netzsch Balance) – Colored Regions used for Analysis